# Structural order matters: Enhanced electronic coupling in self-assembled micro-crystals of Au-nanoclusters

Florian Fetzer[1]*, Andre Maier[2,4]*, Martin Hodas[3], Olympia Geladari[2,4], Kai Braun[2,4], Alfred J. Meixner[2,4], Frank Schreiber[3,4], Andreas Schnepf[1#], Marcus Scheele[2,4#]

*F.F. and A.M. contributed equally to the publication.
#Corresponding authors.

1. Institut für Anorganische Chemie Universität Tübingen, Auf der Morgenstelle 18, D-72076 Tübingen, Germany
2. Institut für Physikalische und Theoretische Chemie, Universität Tübingen, Auf der Morgenstelle 18, D-72076 Tübingen, Germany
3. Institut für Angewandte Physik, Universität Tübingen, Auf der Morgenstelle 10, D-72076 Tübingen, Germany
4. Center for Light-Matter Interaction, Sensors & Analytics LISA+, Universität Tübingen, Auf der Morgenstelle 15, D-72076 Tübingen, Germany

## Abstract

We report an easy and broadly applicable method for the controlled self-assembly of atomically precise $Au_{32}(^nBu_3P)_{12}Cl_8$ nanoclusters into micro-crystals. This enables the determination of emergent optoelectronic properties resulting from long-range order in such assemblies. Compared to the same nanoclusters in glassy, polycrystalline ensembles, we find a 100-fold increase in the electric conductivity and charge carrier mobility as well as additional optical transitions. We show that these effects are due to a vanishing energetic disorder and a drastically reduced activation energy to charge transport in the highly ordered assemblies. This first structure-transport correlation on self-assembled superstructures of atomically precise gold nanoclusters paves the way towards functional materials with novel collective optoelectronic properties.

## Introduction

Using the collective properties of self-assembled molecules and particles as building blocks bears immense opportunities for microelectronic applications.[1–3] Already implemented applications of self-assembled thin films range from light emitting diodes (LED) over field effect transistors (FET) to optical sensors.[4] Inorganic nanoparticles, organic π-systems and conjugated polymers are the most widely used components for such self-assembly.[5–8] For instance, previous studies have shown the possibility to form three-dimensional assemblies with long-range order using gold nanoparticles as building blocks.[9] However, these nanoparticles consist of a few hundred to thousands of atoms, are not atomically precise, exhibit finite size distributions and, thus, an inherent energetic disorder in ensembles. To mitigate this shortcoming, atomically precise, inorganic molecular clusters have been suggested as promising building blocks for customized electronic materials by design of their structure.[10] These materials exhibit larger dielectric constants than organic semiconductors with profound consequences for their excited-state properties, such as the ability to exploit quantum confinement effects. Atomically precise metalloid nanoclusters (NCs) form a subgroup of this material class.[11,12] The exact knowledge of their structure and composition along with usually smaller sizes, enhanced quantum confinement and the prospect of single-electron switching at room temperature promotes NCs as building blocks for self-assembly.[13,14]

Previous studies on Au NC ensembles have either reported conductivity measurements of polycrystalline assemblies,[15] along with the first observation of semiconducting properties,[16] or the formation of highly ordered micro-crystals.[14,17,18] However, attempts to quantify the influence of perfect order on the electronic properties of such micro-crystals have remained unsuccessful.[19] Overcoming this challenge would allow exploiting the distinct properties of perfectly ordered NC micro-crystals, such as superconductance in metalloid $Ga_{84}R_{20}^{4-/3-}$ clusters.[20–22]

In this paper, we show that assemblies of Au$_{32}$($^n$Bu$_3$P)$_{12}$Cl$_8$-nanoclusters form idiomorphic micro-crystals with high crystallographic phase purity and a strongly preferred growth direction. The crystals are semiconducting and exhibit p-type hopping transport which is limited by Coulomb charging. Energetic disorder is negligible in these micro-crystals. In contrast, disordered assemblies of the same clusters show a 100-fold decrease in the electric conductivity and an over 50% larger activation energy for hopping transport due to the disorder.

## Results

**Self-assembly of Au$_{32}$-NC micro-crystals**

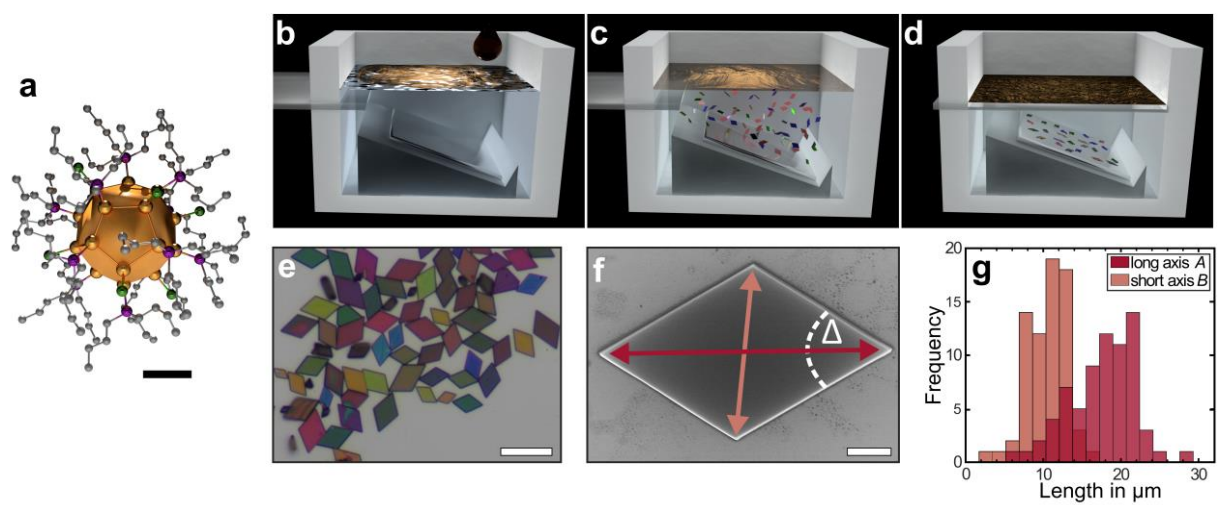

**Figure 1: Au$_{32}$-NC self-assembly into micro-crystals**. (**a**) Structural drawing of an Au$_{32}$($^n$Bu$_3$P)$_{12}$Cl$_8$ NC. The different colors represent the Au- (gold), Cl- (green), P- (purple) and C- (grey) atoms, while hydrogens are omitted for clarity. The Au-core has a diameter of ~0.9 nm, while the size of the entire NC is about 1.3 nm. Scale bar: 0.4 nm. (**b-d**) Schematic illustration of the assembly process. Details are given in the Methods section. (**e**) Optical micrograph of self-assembled Au$_{32}$-NC micro-crystals on a Si/SiO$_x$ substrate. The crystals are µm-sized and exhibit a parallelogram shape. Different sizes and thicknesses (color) can be observed. Scale bar: 15 µm. (**f**) SEM micrograph of a self-assembled micro-crystal with indicated long axis $A$, short axis $B$ and angle $\Delta$. Scale bar: 2 µm. (**g**) Distribution of long and short axis revealing typical crystal sizes of $A = 17.4 \pm 4.2$ µm, $B = 10.6 \pm 2.5$ µm (polydispersity of 24%).

The atomically precise building blocks of metalloid $Au_{32}(^{n}Bu_3P)_{12}Cl_8$ nanoclusters (abbreviated as $Au_{32}$-NCs) with an Au-core size of ~0.9 nm are synthesized as previously described.[23] Including the full ligand shell of twelve phosphine ligands and eight chloride atoms, the building block size is about 1.3 nm, displayed in **Figure 1a**. Single crystal X-ray diffraction of macroscopic crystals of $Au_{32}$-NCs yields a triclinic unit cell containing two crystallographically independent NCs ($a = 1.91$ nm, $b = 1.93$ nm, $c = 3.32$ nm; $\alpha = 73.2°, \beta = 86.7°, \gamma = 63.4°$, space group $P\bar{1}$).[23]

The preparation process, where dispersed $Au_{32}$-NCs self-assemble into micro-crystals at the liquid-air interface and sink into the liquid subphase, is schematically illustrated in **Figure 1b-d**. This method allows the preparation of micro-crystals onto any substrate of interest (Further details on the preparation can be found in Methods and Supplementary Information). By 'micro-crystals' we understand micrometer-sized idiomorphic single crystals of $Au_{32}$-NCs with high crystallographic phase purity and a strongly preferred growth direction, as detailed below. **Figure 1e** shows a typical ensemble of self-assembled $Au_{32}$-NC micro-crystals with parallelogram shape on a silicon wafer. The crystal shape can be quantified by its geometrical properties of long axis $A$, short axis $B$, angle at the sharp edge $\Delta$ and thickness $h$, as illustrated in the SEM micrograph in **Figure 1f**. The lateral expansion (5–30 µm) is 2–3 orders of magnitude larger than the thickness (50–600 nm, see Supplementary **Figure S1**), indicating a strongly preferred growth direction. An analysis of SEM micrographs of individual micro-crystals yields a distribution of long and short axis, revealing a typical lateral size of $A = 17.4 \pm 4.2$ µm and $B = 10.6 \pm 2.5$ µm, as indicated in **Figure 1g**. The lateral size dispersion is calculated to 24%. Further, we observe an aspect ratio of long and short axis of $A/B = 1.64$ and a sharp edge angle of $\Delta = 63°$ for all micro-crystals. This aspect ratio corresponds to the associated ratio found in the unit cell of macroscopic $Au_{32}$-NC crystals, and the angle $\Delta$ suits the $\gamma$-angle of the unit cell of $\gamma = 63.4°$.[23] Hence, the shape of the micro-crystals strongly resembles the aforementioned unit cell which renders the crystals idiomorphic. High resolution

SEM images (see Supplementary **Figure S1**) reveal perfectly defined edges and extremely flat surfaces, indicating a high crystalline phase purity. Different color impressions in **Figure 1e** originate from interference phenomena indicating different thicknesses.

**Structural investigation of self-assembled Au$_{32}$-NC micro-crystals**

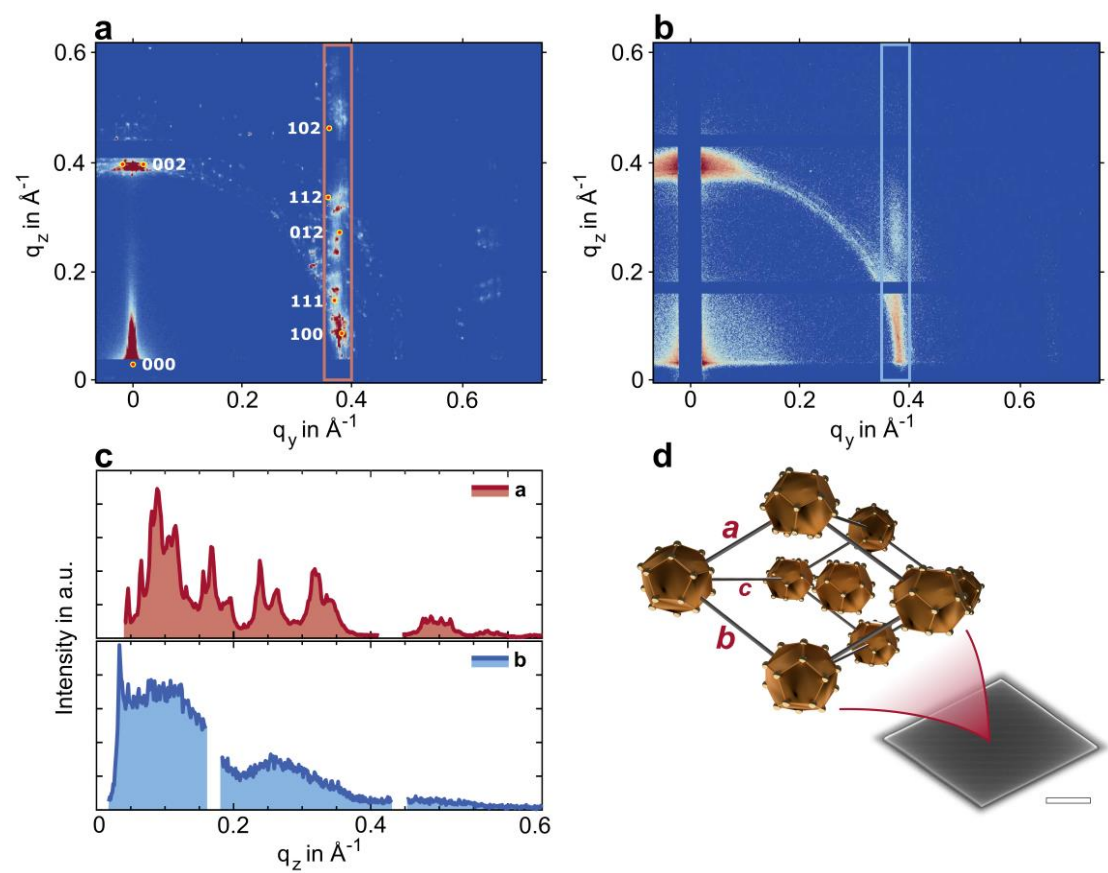

**Figure 2: Structure of Au$_{32}$-NC micro-crystals**. (**a**) GISAXS pattern of an ensemble of hundreds of micro-crystals with different azimuthal orientation. Diffraction spots are simulated according to a triclinic unit cell ($a$ = 1.90 nm, $b$ = 1.94 nm, $c$ = 3.48 nm and $\alpha$ = 72°, $\beta$ = 86°, $\gamma$ = 59°). (**b**) GISAXS pattern of a spin-coated thin film of Au$_{32}$-NCs. (**c**) Line scans along $q_z$ at $q_y$ = 0.37 Å$^{-1}$ of the pattern in red (a) and blue (b), respectively, highlighted by the rectangular boxes. The ensemble of micro-crystals (a) show distinct sharp peaks, indicating the high crystallinity. The polycrystalline sample (b) shows broad signals while lacking clear peaks, indicating the polycrystalline and defect-rich structure. Gaps correspond to detector edges. (**d**) Schematic drawing of the triclinic unit cell with axis $a$, $b$ and $c$ indicating the idiomorphic growth of the displayed micro-crystal. The unit cell contains two crystallographically

independent NCs. Ligand spheres are omitted for clarity. The scale bar of the SEM micrograph of a micro-crystal corresponds to 3 µm.

To verify the crystallinity of self-assembled micro-crystals, grazing-incidence small-angle X-ray scattering (GISAXS) measurements are performed, which is a common technique to investigate the structural properties of nanoparticle assemblies in thin films or at interfaces.[24–26] The GISAXS pattern of an ensemble of hundreds of individual micro-crystals with different azimuthal orientation (**Figure 1e**) is shown in **Figure 2a**. Sharp peaks are obtained (**Figure 2c**), indicating the high crystallinity of the sample. Doubled peaks in *z* direction can be observed, caused by a peak splitting phenomenon as previously described.[27] The fit of the obtained peaks yields a triclinic unit cell ($a$ = 1.9 nm, $b$ = 1.94 nm, $c$ = 3.48 nm and $\alpha$ = 72°, $\beta$ = 86°, $\gamma$ = 59°), which is simulated onto the diffraction pattern. The fit is in excellent agreement with the previously determined unit cell of a macroscopic $Au_{32}$-NC single crystal ($a$ = 1.91 nm, $b$ = 1.93 nm, $c$ = 3.32 nm and $\alpha$ = 73.2°, $\beta$ = 86.7°, $\gamma$ = 63.4°).[26] Considering the GISAXS data together with the morphological appearance of self-assembled $Au_{32}$-NC micro-crystals, the unit cell of the micro-crystals can be described by a triclinic structure with axis ratios and angles corresponding to a macroscopic single crystal of $Au_{32}$ (**Figure 2d**). Thus, micro-crystals are µm-sized single crystals, built from individual building blocks of $Au_{32}$-NCs. A typical micro-crystal consist of ~5,000 unit-cells laterally along the long axis *A* and ~15–200 unit-cells out-of-plane (~$10^9$ $Au_{32}$-NCs per micro-crystal). Furthermore, the dominant first peak in *z* direction at $q_z \approx 0.37$ Å$^{-1}$ corresponds to a distance of about $d$ = 1.7 nm. Assuming this to be the {002}-peak (based on the bulk structure), a unit cell edge of 3.4 nm can be calculated which is in good agreement with the unit cell length $c$ = 3.32 nm of the macroscopic NC crystal, indicating that the *c* axis of the unit cell is aligned along the surface normal. In combination with the missing peaks at {200} and {020}, we conclude that most micro-crystals lay flat on the substrate surface, with axis *a* and *b* oriented parallel to the substrate, as it is observed by microscopy techniques (Supplementary **Figure S2**). Some peaks along the ring-like features at $q \approx 0.37$ Å$^{-1}$ are observed and attributed to single-crystals which are not oriented flat on the surface and residual agglomerations which are not Au$_{32}$-NC micro-crystals (see Supplementary **Figure S2**). In comparison to micro-crystals, the GISAXS pattern of a spin-coated $30 \pm 2$ nm thin film is given in **Figure 2b**. Instead of sharp peaks, more ring-like and smeared peaks are observed, clearly indicating the polycrystalline and defect-rich structure of the sample. Throughout this work we refer to these samples as 'polycrystalline' to indicate their low degree of crystallinity and high angular disorder.

**Optical properties of Au$_{32}$-NC micro-crystals**

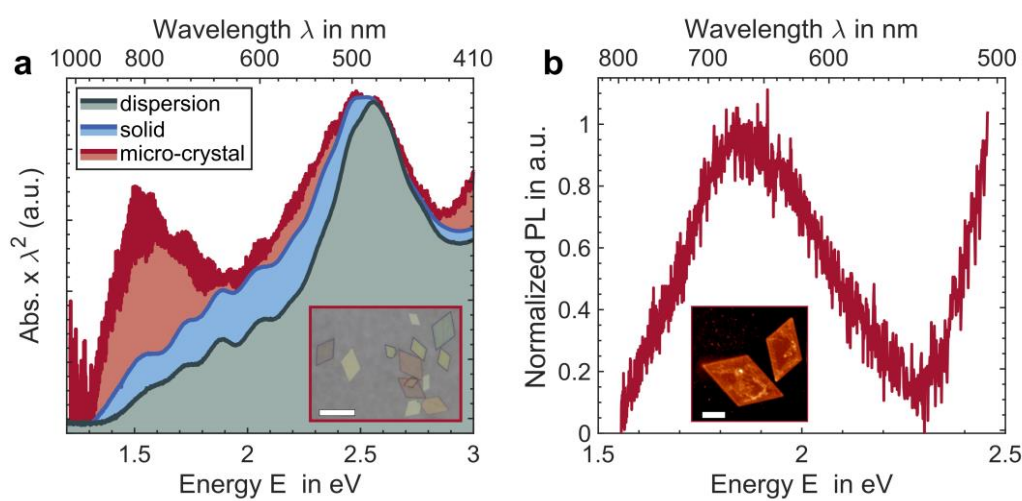

**Figure 3**: **Optical properties of Au$_{32}$-NCs and micro-crystals**. (**a**) Absorbance spectra of Au$_{32}$-NCs dispersed in hexane (green), in a thin film (blue) and in micro-crystals (red) on glass. Individual micro-crystals show enhanced absorption at 800 nm (1.55 eV), corresponding to the HOMO-LUMO transition. The prominent absorption peak at 481 nm (2.58 eV) for dispersed NCs, is slightly redshifted to 483 nm (2.51 eV) and broadened for Au$_{32}$-NCs in thin films and micro-crystals. All spectra are normalized to the prominent peak at 481 nm. The inset shows an optical micrograph of individual micro-crystals on a glass substrate. Scale bar: 15 µm. (**b**) Photoluminescence (PL) spectrum of an individual micro-crystal shows a broad emission peak at around 670 nm. The inset displays the luminescence of two micro-crystals upon excitation with 488 nm. Scale bar: 5 µm.

The comprehensive characterization of the micro-crystals is concluded by optical and electronic investigations. **Figure 3a** displays the energy-corrected absorbance spectra of a $Au_{32}$-NCs dispersion, a thin film and a micro-crystal. Dispersed $Au_{32}$-NCs in solution exhibit several distinct peaks and shoulders, attributed to molecular-like transitions (full spectrum in Supplementary **Figure S3**). While the most prominent absorption peak is observed at 2.58 eV (481 nm), the first absorption peak at 1.55 eV (800 nm) corresponds to the HOMO-LUMO transition.[28,29] Most strikingly, only in micro-crystals of $Au_{32}$-NCs this peak is strongly enhanced as shown in **Figure 3a**. Further, the absorption onset as well as the most prominent peak at 2.57 eV are red-shifted by approximately 100 meV and 10 meV, respectively. A generally enhanced absorption at lower energies and a broadening/shoulder formation at 2.48 eV (500 nm) are observed in micro-crystals and thin films of $Au_{32}$-NCs. We attribute these findings to a gradual progression from virtually no electronic coupling between the $Au_{32}$-NCs in solution to weak coupling in thin films and enhanced electronic interactions in the highly ordered micro-crystals.[2,19,30,31]

While no emission of the $Au_{32}$-NCs is observed in solution, $Au_{32}$-NC micro-crystals exhibit photoluminescence resulting in a broad emission peak at 670 nm (1.85 eV) after excitation at $\lambda_{ex}$ = 488 nm, as shown in **Figure 3b**.

**Electronic properties of Au$_{32}$-NC micro-crystals**

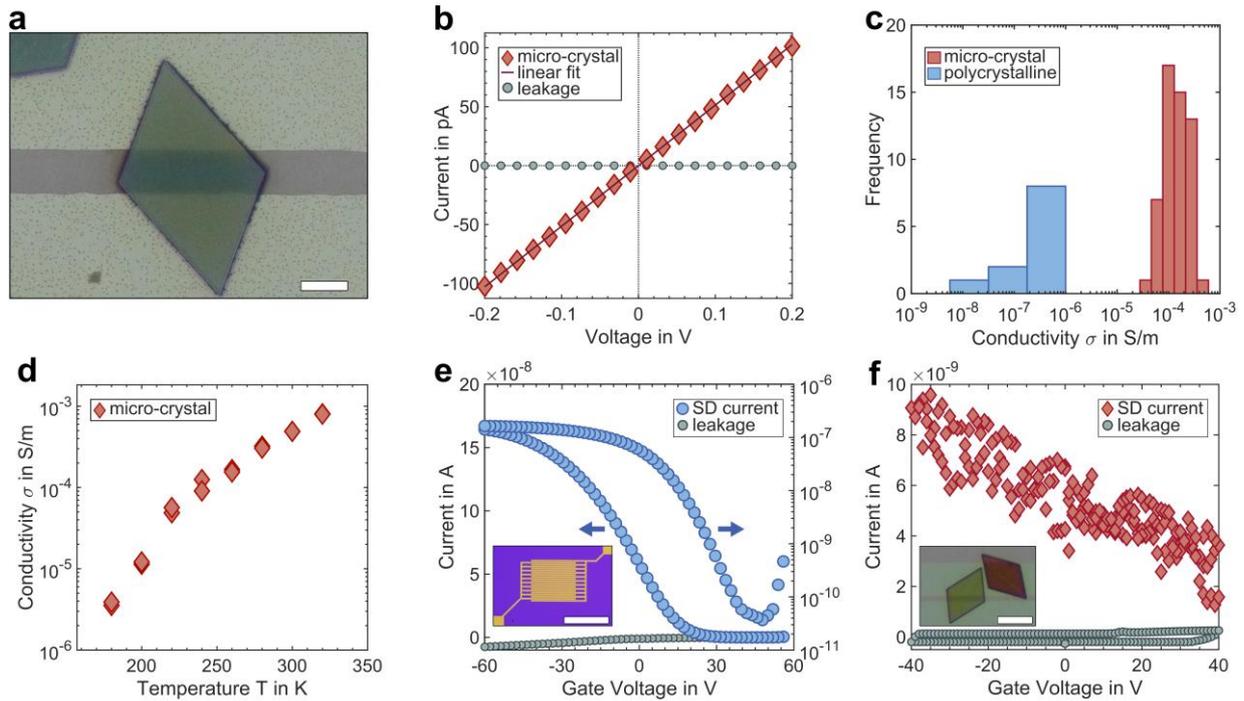

**Figure 4: Electronic properties of Au$_{32}$-NC micro-crystals**. (**a**) SEM micrograph of an individual micro-crystal deposited on two Au-electrodes on a Si/SiO$_x$ device. The electrodes form a channel with length $L = 2.8$ µm. The width and height of the contacted micro-crystal are $W = 7.9 \pm 0.4$ µm and $h = 120$ nm. SEM and optical micrographs are merged. (**b**) Typical *I-V* curve of an individually probed micro-crystal. Ohmic behavior is observed in the low voltage regime. (**c**) Distribution of conductivity $\sigma$ of 54 individual micro-crystals and 19 polycrystalline thin films. The conductivity of micro-crystals exceeds that of polycrystalline films by ~2 oders of magnitude. (**d**) Temperature-dependent conductivity of Au$_{32}$-NC micro-crystals with two individual measurements per temperature step. (**e**) FET transfer curve (blue) of a polycrystalline film of Au$_{32}$-NCs on an interdigitated electrode device with $L = 2.5$ µm, $W = 1$ cm, measured at $V_{SD} = 10$ V on a linear and logarithmic scale together with the negligible leak current (grey). Arrows indicate the corresponding *y*-axis. (**f**) FET transfer curve (red) of an individual micro-crystal device with $L = 1.5$ µm, $W = 10.4 \pm 0.2$ µm, measured at $V_{SD} = 5$ V, together with the negligible leak current (grey). The insets in (e) and (f) display optical micrographs of the two devices. Scale bars correspond to 500 µm and 10 µm, respectively.

To study the possible electronic coupling between individual Au$_{32}$-NCs observed via optical spectroscopy, we perform (temperature-dependent) conductivity and field-effect transistor (FET) measurements on single Au$_{32}$-NC micro-crystals. Most remarkably, we find that the conductivity of highly ordered Au$_{32}$-NCs within micro-crystals exceeds that of polycrystalline assemblies by two orders of magnitude, corroborating our hypothesis of enhanced electronic coupling.

We designed electrode devices, in which deposited micro-crystals bridge adjacent electrodes to be addressed and probed individually. Details on the device layout are given in the Supplementary Information (Supplementary **Figure S4** and **S5**). **Figure 4a** shows an SEM micrograph of a 120 nm thick micro-crystal deposited on two Au-electrodes with a gap of $L = 2.8$ µm on a Si/SiO$_x$ device. **Figure 4b** displays a typical *I-V* curve of an individual micro-crystal in the range of ±200 mV. Ohmic behavior (at room-temperature) in the low-field regime (up to ±1 V) is observed. Electrical conductivity values with typical uncertainties of <10% are calculated from these measurements for 54 individual micro-crystal channels on different devices. **Figure 4c** displays the narrow distribution of conductivity values, showing a mean conductivity of $\sigma = 1.56 \times 10^{-4}$ S/m with a standard deviation of $\pm 0.90 \times 10^{-4}$ S/m. In contrast, the mean conductivity of polycrystalline thin films of Au$_{32}$-NCs is only $\sigma \approx 1 \times 10^{-6}$ S/m. These devices are obtained by spin-coating on substrates with interdigitated electrodes of channel length $L = 2.5$ µm and width $W = 1$ cm. The film thicknesses are in the range of 30 ± 2 nm to 47 ± 4 nm.

To shed light on the charge-transfer mechanism of electronic transport within micro-crystals and polycrystalline films of Au$_{32}$-NCs, we perform temperature-dependent conductivity measurements at $T = 340$–$170$ K (**Figure 4d**). Below this range, the measured current approaches the noise level. The measured temperature dependence can be described by an Arrhenius-type temperature-activated hopping (Supplementary **Figure S6**).[32] Fitting the conductivity data accordingly, we obtain activation energies of $E_A = 227 \pm 17$ meV for individual micro-crystals and $E_A = 366 \pm 62$ meV for the spin-coated polycrystalline Au$_{32}$-NCs thin films.

To further characterize the electronic properties of self-assembled $Au_{32}$-NC micro-crystals and polycrystalline films, FET measurements are performed. Strikingly, a field-effect can be observed, indicating semiconducting behavior of the metal NC assemblies. **Figure 4e** shows the FET transfer curve of a polycrystalline thin film of $Au_{32}$-NCs on interdigitated electrodes with channel dimensions of $L = 2.5$ µm and $W = 1$ cm. p-type behavior is observed, indicating holes ($h^+$) as majority charge carriers. The current flow can be modulated by more than three orders of magnitude. Ambipolar behavior is also observed for very high threshold voltages of $V_G > 40$ V. The calculated hole mobility of spin-coated $Au_{32}$-NC films is in the range of $\mu(h^+) \sim 10^6 - 10^5$ cm$^2$ V$^{-1}$ s$^{-1}$.

**Figure 4f** displays the FET transfer curve of an individual $Au_{32}$-NC micro-crystal, which also indicates p-type behavior. Note that the current flows through a much smaller channel width $W$ of 5–10 µm in this case. Here, the mean value of the hole mobility can be calculated to be $\mu(h^+) = 0.8 \times 10^{-4}$ cm$^2$ V$^{-1}$ s$^{-1}$. The noise in current flow and the low modulation can be attributed to the non-ideal channel geometry. Further, the quality of contact between the dielectric $SiO_x$ layer and the micro-crystal is not known. The non-ideal contact might influence the appearance of transfer curves.

Knowing the charge carrier mobility $\mu(h^+)$ and the conductivity of individual $Au_{32}$-NC crystals, we calculate the charge carrier concentration to be $n(h^+) = 2 \times 10^{17}$ cm$^{-3}$. This correspond to one free charge carrier per 1000 $Au_{32}$-NCs, as the concentration of individual $Au_{32}$-NC within a crystal is $1.9 \times 10^{20}$ cm$^{-3}$.

## Discussion

The Au$_{32}$-NC HOMO-LUMO gap of 1.55 eV (**Figure 3a**) is consistent with earlier reports on other Au NCs and the expected degree of quantum confinement. Specifically, for NCs with 11 and 25 Au atoms and, thus, stronger quantum confinement, HOMO-LUMO transitions of 2.97 eV and 1.84 eV have been reported.[16,33] A related size-dependent study of NCs with 10 to 39 Au core atoms revealed HOMO-LUMO transitions from 3.7 eV to 1.7 eV.[34]

The solid state luminescence of the Au$_{32}$-NCs (**Figure 3b**) at 1.85 eV is fully consistent with the emission of other Au-NCs,[11,34–37] such as Au$_{25}$,[28,38–40] and may be attributed to aggregation-induced emission.[39,41,42] In contrast to the HOMO-LUMO transition, which is believed to involve a (mostly dark) sp-intraband transition, the luminescence in Au$_{25}$ and Au$_{28}$-NCs results from an sp→d interband transition, which may also be the case in Au$_{32}$.[37,38] We note, however, that Au$_{25}$-NCs consist of an icosahedral Au$_{13}$ core, while the core of the Au$_{32}$-NC is a hollow Au$_{12}$ icosahedron with potentially different optical properties.[11,23,43]

The conductivity (**Figure 4c**) and mobility (**Figure 4e**) of the thin polycrystalline Au$_{32}$-NC films are in good agreement with previously reported values for Au$_{25}$- and Au$_{38}$-NCs.[16,44] In contrast to the study by Galchenko et al. on Au$_{25}$-NCs with n-type transport,[16] we observe here p-type behavior or ambipolar transport with extremely high threshold voltages of approximately $V_G = +50$ V.

The key finding of this work is that the above-mentioned properties change dramatically as long-range order is introduced to the NC ensembles (**Figure 2a**). To the best of our knowledge, no transport studies on single-crystalline domains of atomically precise Au NCs have been achieved before and, thus, this is the first time that the effect of structural coherence on transport can be quantified.

To this end, we use the experimentally determined activation energies to charge transport in the Au$_{32}$-NCs, either as micro-crystals ($E_A = 227 \pm 17$ meV) or as polycrystalline thin films ($E_A = 366 \pm 62$ meV). Transport in weakly coupled nanostructures depends on the transfer integral ($\delta$), the Coulomb charging energy ($E_C$) and the energetic disorder $\Delta\alpha$.[45] Strongly temperature-activated transport (Figure 4d) suggests that even the Au$_{32}$-NC micro-crystals are in the Mott regime with $E_C \gg \delta$. Thus, charge transport is dominated by $E_C$ and possibly $\Delta\alpha$. $E_C$ can be referred to as the self-capacitance of the NC and it describes the required energy for addition or removal of an additional charge carrier to the NC. We estimate $E_C$ of the micro-crystals to 276 meV (for details, see Supplementary Information), which is consistent with the full activation energy. Thus, charge carrier transport in the micro-crystals depends solely on the charging energy and the energetic disorder is negligible. In contrast, $E_A$ in the polycrystalline thin films largely exceeds $E_C$, suggesting a significant degree of energetic disorder, which is caused by structural, orientational or chemical disorder of the individual NCs. Since the NCs are atomically precise, we hold only structural defects, such as grain boundaries, cracks and a lack of orientational order to be responsible for the occurrence of a non-zero $\Delta\alpha$ in the polycrystalline films.[46] This effect is especially pronounced here, as systems with large $E_C$ are generally very sensitive towards structural disorder.[17] In contrast, Au$_{32}$-NC micro-crystals not only consist of chemically identical building blocks but also exhibit structural perfection, which manifests in a vanishing value of $\Delta\alpha$. We suggest that this is the reason for the enhanced electronic coupling and altered optoelectronic properties. Future attempts to further increase coupling in Au NC micro-crystals should focus on increasing the transfer integral, for instance by reducing the distance between adjacent clusters or by covalent coupling with conjugated linkers. If $\delta \approx E_C$, a Mott insulator-metal transition occurs and band-like transport becomes possible. The basis for this will be atomically defined building blocks in combination with a suitable coupling as pioneered here.

## Conclusion

Atomically precise Au$_{32}$($^n$Bu$_3$P)$_{12}$Cl$_8$ nanoclusters are self-assembled into micro-crystals with high crystallographic phase purity and a strongly preferred growth direction. Individual micro-crystals exhibit semiconducting p-type behavior and temperature activated hopping transport, limited by Coulomb charging. Most strikingly, additional optical transitions emerge, and electronic coupling is 100-fold enhanced in the micro-crystals compared to polycrystalline thin films, highlighting the advantageous effect of long-range structural order. This study implies that utilizing atomically precise building blocks for the self-assembly into superlattices eliminates energetic disorder and provides a promising route towards self-assembled nanostructures with emergent optoelectronic properties.

## Acknowledgements

F.F. thanks the PhD Network: "Novel nanoparticles: from synthesis to biological applications" at the University of Tübingen for financial support. This project has been funded by the European Research Council (ERC) under the European Union's Horizon 2020 research and innovation program (grant agreement No 802822). M.H. thanks the Alexander von Humboldt foundation for financial support. SEM measurements using a Hitachi SU 8030 SEM were funded by the DFG under contract INST 37/829-1 FUGG. Support by C. Dreser, S. Dickreuter and A. Bräuer with the optical setup of the M. Fleischer group is gratefully acknowledged. We thank B. Fischer for his support during the interference reflection microscopy measurements.

## Authors contribution

F.F. synthesized the NCs and developed the micro-crystals fabrication. A.M. performed the device fabrication, SEM measurements and analyzed the electrical measurements. F.F. and A.M. conducted optical absorbance measurements, electrical measurements, interpreted the results and wrote the manuscript. M.H. performed the GISAXS measurements and analysis. O.G., K.B. and A.J.M. conducted the luminescence measurements. F.S., A.S. and M.S. conceived and supervised the project. All authors have given approval to the final version of the manuscript.

## Competing interests

The authors declare no competing interests.

# Methods

**Materials:** All chemicals were used as received unless otherwise noted. $Au_{32}(^nBu_3P)_{12}Cl_8$-nanoclusters were synthesized as described elsewhere.[1] Octane and acetonitrile were bought from Sigma-Aldrich and were degassed and distilled before usage. Silicon/silicon dioxide ($Si/SiO_x$) wafer with 200 nm $SiO_x$ layer and n-doped Si were purchased from Siegert Wafer. Photoresist, developer and remover (ma-N 405, ma-D 331/S and mr-Rem660, respectively) were purchased from micro resist technology GmbH, Berlin.

**Self-assembly of $Au_{32}$-NC micro-crystals:** The formation of crystals via liquid-air interface method is schematically illustrated in Figure 1b-d. A solution of $Au_{32}$-NCs in octane (200 µl, 0.5 mM) was added onto a subphase of acetonitrile inside a home-built Teflon chamber (Figure 1b). The self-assembly of $Au_{32}$-NCs into micro-crystals took place at the phase boundary between the acetonitrile subphase and the NC solution upon evaporation of the solvent. The micro-crystals started to sink down through the subphase and stuck to the desired substrate which was previously placed inside the liquid subphase (Figure 1c). After 45 min a glass slide was horizontally inserted into the subphase to separate the residual $Au_{32}$-membrane (floating on the liquid-air interface) from the bottom substrate (Figure 1d). The liquid subphase was removed and the substrate dried at ambient conditions. Micro-crystal fabrication took place at ambient condition. Further details are given in the Supporting Information and Figure S7.

**Device fabrication:** For the micro-crystal electrode devices, standard photolithography technique (negative tone resist) was used to pattern Au electrodes on $Si/SiO_x$ substrates (200 nm $SiO_x$). Au (8–10 nm) and Ti (~2.5 nm) as an adhesion layer were thermally evaporated under high vacuum conditions. Ultrasonic-assisted lift-off in mr-Rem 660 removed the residual resist and metal layer. Electrodes with gaps of 1.5-2.5 µm (channel length *L*) were realized. Devices were coated with micro-crystals as described above and checked with a basic light microscope to identify channels, where a single micro-crystal bridges two adjacent electrodes. For thin film devices with interdigitated electrodes, commercially available OFET substrates (Fraunhofer IPMS, Dresden) were purchased. The substrates were coated with 100–200 µl of a 0.5 mM $Au_{32}$-NC solution (hexane or heptane) and spin-coated after 2 min with a speed of 2000 rpm for 30 s. All devices were prepared at ambient conditions in a fume-hood.

**Grazing-incidence small-angle X-ray scattering:** GISAXS measurements were conducted on a Xeuss 2.0 setup (Xenocs). A CuK$\alpha$ X-ray beam with wavelength $\lambda = 1.5418$ Å ($E = 8.04$ keV) and a beam size of ~ 500 × 500 µm² (FWHM) was used. A two-dimensional detector Pilatus 300K (Dectris) with 487 × 619 pixels of 175 × 175 µm² was positioned 365 mm downstream the sample. The samples (micro-crystal ensemble or thin film on Si wafer with 200 nm $SiO_x$ layer) were probed at an incidence angle of 0.2°. Acquisition times of 60 min and 30 min were used to obtain the GISAXS maps in Figure 2a and Figure 2b, respectively. Simulated peaks correspond to a triclinic unit cell with $a = 1.90$ nm, $b = 1.94$ nm, $c = 3.48$ nm and $\alpha = 72°$, $\beta = 86°$, $\gamma = 59°$, which is in good agreement with the X-ray diffraction data from macroscopic $Au_{32}$ NC crystals ($a = 1.91$ nm, $b = 1.93$ nm, $c = 3.32$ nm and $\alpha = 73.2°$, $\beta = 86.7°$, $\gamma = 63.4°$). Simulations were performed using the MATLAB toolbox GIXSGUI.[2]

**Optical measurements:** Absorbance spectra of $Au_{32}$-NC in solutions (0.5 mM in hexane) were acquired with an UV-vis-NIR spectrometer (Cary 5000, Agilent Technologies). For thin films spin-coated on glass slides, a Perkin Elmer Lambda 950 spectrometer was used. For individual micro-crystals on glass slides, an inverted microscope (Nikon Eclipse Ti-S) with a spectrometer was used. The sample was illuminated with unpolarized white light by a 100 W halogen lamp. The transmitted light was collected by a 60× objective (Nikon, CFI S Plan Fluor ELWD, NA = 0.7). The collected light was passed to a grating spectrograph (Andor Technology,

Shamrock SR-303i) and detected with a camera (Andor Technology, iDusCCD). All absorbance spectra were energy corrected using the expression $I(E) = I(\lambda) \times \lambda^2$.[3,4] Photoluminescence images and emission spectra of individual $Au_{32}$-NC micro-crystals were acquired with a home-built confocal laser scanning microscope. The diode laser (iBeam smart, Toptica Photonics) was operated in continuous wave Gaussian mode at an excitation wavelength of $\lambda_{ex}$ = 488 nm. Luminescence images were obtained with a photon counting module (SPCM-AQR-14, PerkinElmer) and spectra were acquired with an UV-VIS spectrometer (Acton SpectraPro 2300, Princeton Instruments). The background was subsequently subtracted from the emission spectra.

**Scanning electron microscopy:** SEM imaging of micro-crystals on Si/SiO$_x$ devices was performed with a HITACHI model SU8030 at 30 kV. To estimate the thickness of micro-crystals, samples were titled by 85° with respect to the incoming electron beam.

**Electrical measurements:** All electrical measurements were conducted under vacuum in a probe station (Lake Shore, CRX-6.5K). All samples were placed under vacuum over night before measurement (pressure of <10$^{-5}$ mbar). Au-electrode pairs were contacted with W-tips, connected to a source-meter-unit (Keithley, 2636 B). A back electrode worked as gate electrode. For two-point conductivity measurements, voltage sweeps in a certain range of ±1 V were applied and the current (as well as leak current) detected. Fitting the linear *I-V* curve yielded the conductance value *G*. Conductivity $\sigma$ was calculated as $\sigma = (G \times L)/(W \times h)$. The dimensions length, width, thickness (*L*, *W*, *h*) were determined by SEM imaging for micro-crystals or profilometry for spin-coated thin films (Dektak XT-A, Bruker). For FET measurements (bottom-gate, bottom-contact configuration), a source-drain voltage of $V_{SD}$ was applied and $I_{SD}$ was measured, modulated by applied gate voltages $V_G$. Using the gradual channel approximation, field effect mobilities $\mu$ were calculated (Supplementary Equation S1).

For temperature-dependent measurements, the devices were cooled down to 8 K and gradually heated with a Lake Shore temperature controller (model 336). Current was detected in the temperature range 170–340 K. At least two measurements were taken for every temperature. After reaching 340 K, measurements were repeated at lower temperature, to verify the reversibility. The temperature-activated hopping behavior can be described as an Arrhenius-type, which is expressed in Equation 1.[5]

$$\sigma = \sigma_0 \exp(-E_A/k_B T) \qquad (1)$$

Here, $E_A$ is the activation energy, $k_B$ the Boltzmann constant, $T$ the temperature and $\sigma_0$ a constant. $E_A$ was obtained from the slope of $\ln(\sigma)$ as a function of $T^{-1}$.

**Method References**

# – Supplementary Information–

**SEM imaging of micro-crystals**

Figure S1a displays a scanning electron micrograph of a micro-crystal. Well-defined edges as well as an extremely flat surface are observed. Figures S1b-d show side-views of micro-crystals with different thicknesses. In side-view the sample is tilted by 85° with respect to the incoming electron beam. From this, the thickness of individual micro-crystals can be investigated.

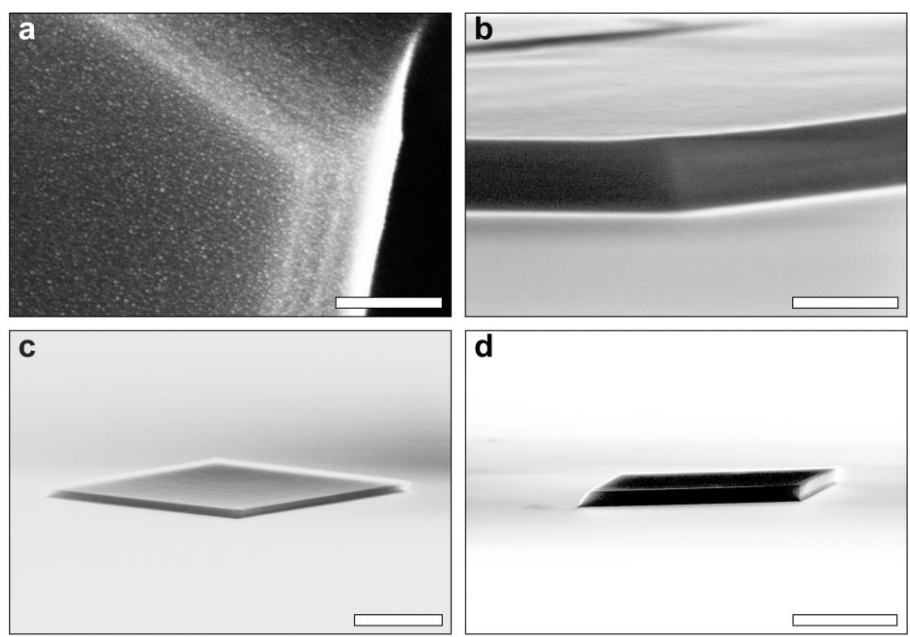

**Figure S1**: **SEM investigation of Au$_{32}$-NC micro-crystals.** (**a**) High-resolution SEM micrograph of an edge of a microcrystal. Individual NCs can be identified on the top-most layer as well as on the slightly tilted sidewall. However, structural arrangement of NCs cannot be resolved. Scale bar: 100 nm. (**b-d**) Side-view of different micro-crystals with thicknesses of ~200 nm (**b**), ~100 nm (**c**) and ~450 nm (**d**) under incident angle of 85°. Scale bars correspond to 300 nm (**b**), 1 µm (**c)** and 3 µm (d).

## Micro-crystal sample for GISAXS measurements

Figure S2 shows an optical micrograph of the micro-crystal sample used for GISAXS. The majority of micro-crystals is oriented flat on the surface. Minor agglomerates cause distortions of the resulting GISAXS pattern, as described in the main text.

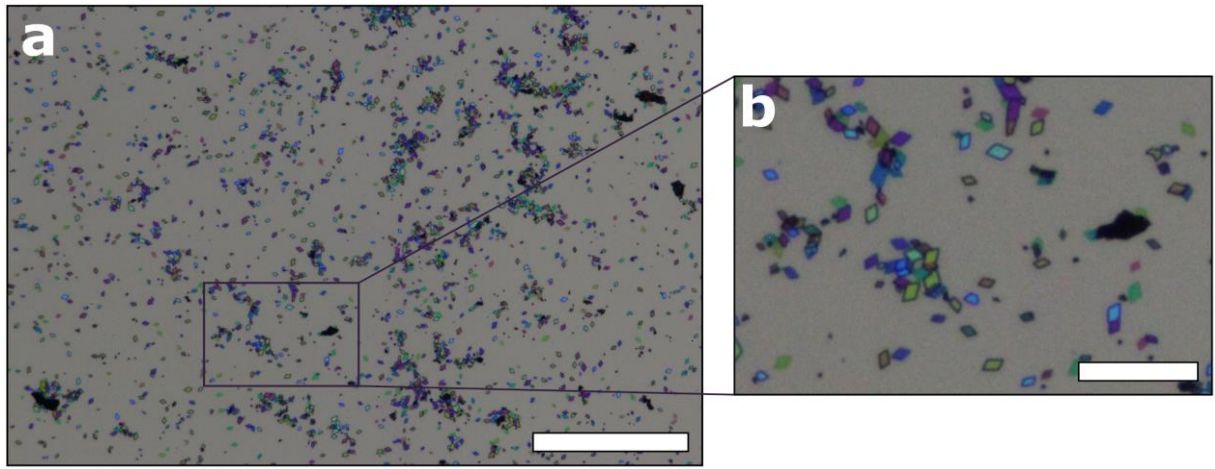

**Figure S2**: Optical micrograph of a Si/SiO$_x$ sample with an ensemble of hundreds of individual micro-crystals used for GISAXS measurements. Scale bars in (a) and (b) correspond to 200 µm and 50 µm, respectively.

# Optical spectrum of Au$_{32}$($^n$Bu$_3$P)$_{12}$Cl$_8$ nanoclusters

Figure S3 shows an optical absorbance spectrum of Au$_{32}$-NCs dissolved in hexane. Several distinct molecular-like transitions are observed, together with a HOMO-LUMO transition at 1.55 eV.

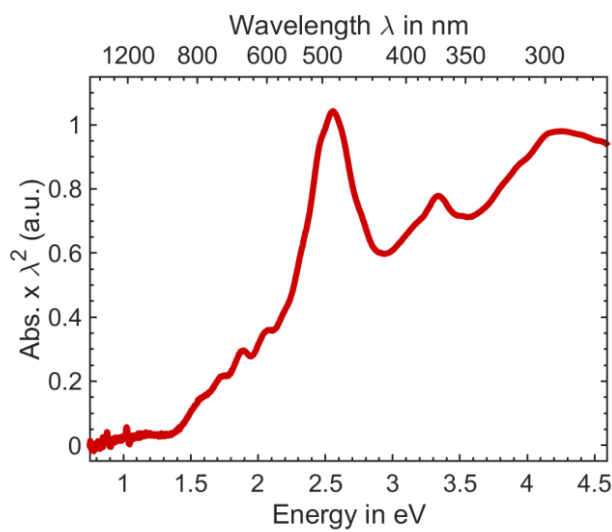

**Figure S3 Optical spectrum of Au$_{32}$-NCs**. Absorbance spectra of Au$_{32}$-NCs dispersed in hexane. Absorbance is energy corrected using the expression $I(E) = I(\lambda) \times \lambda^2$. The peak at 1.55 eV (800 nm) is attributed to the HOMO-LUMO transition. This spectrum corresponds to the spectrum shown in Figure 3a with larger energy range.

# Electrode layout for electronic measurements of individual micro-crystals

Optical micrographs of typical electrode devices (Si/SiO$_x$ substrate) are given in Figure S4. Adjacent electrodes form channels due to overlapping ends. The width of the overlap is 80 µm and the distance between electrodes defines the channel length *L*. On a single device, up to 330 individual channels are realised. By contacting the contact-pads of adjacent electrodes, every channel can be addressed individually. The devices are coated with micro-crystals as described in the Methods section. Micro-crystals which are bridging two adjacent electrodes can be contacted and probed individually. On a typical device, 10–40 individual micro-crystals can be investigated. Due to the relatively thin electrode thickness of ~10 nm, contacted micro-crystals are not free-standing but establish contact to the SiO$_x$ layer, manifested by the observed field-effect.

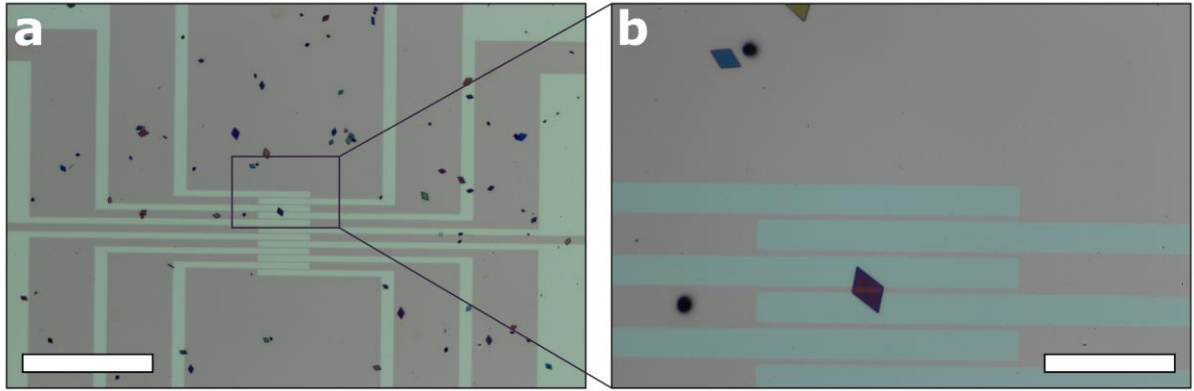

**Figure S4:** Optical micrographs of a typical electrode device with micro-crystals. Micro-crystals which are bridging two adjacent electrodes can be contacted and probed individually. (10 nm Au-electrodes on Si/SiO$_x$ wafer with 200 nm oxide thickness). Scale bars of (a) and (b) correspond to 200 µm and 40 µm, respectively.

**Evaluating the effective width of a micro-crystal within a channel**

The effective width of a micro-crystal on a channel is described best by the mean of $W$ along the channel. Figure S5 shows a SEM micrograph of a micro-crystal covering two channels. The measured conductance $G$ of the channels is different, due to differences in effective width. For every channel, different widths are present (caused by the parallelogram shape). The calculated conductivities $\sigma$ of the two channels should be the same, since the same micro-crystal is probed. Normalizing the measured conductance $G$ with the channel geometry $L/W$ gives essentially the same value (as thickness $h$ is the same). This geometry normalized conductance values, using the mean width, are identical with 39.5 pS and 39.9 pS for the micro-crystals in the two different channels, respectively. Thus, using the mean width along the electric field is the most appropriate dimension.

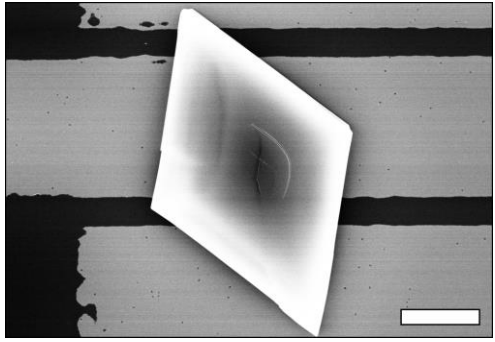

**Figure S5:** SEM micrograph of an individual micro-crystal, covering two different channels. Using the mean value of *W* to calculate the conductivity $\sigma$ yields the same value for geometry normalised conductance for both channels bridged by the same micro-crystal. Scale bar: 5 µm.

## Equations used for the investigation of electronic properties

The field-effect mobilities $\mu$ of individual micro-crystals or polycrystalline thin films are calculated using the gradual channel approximation, given in Equation S1. The charge carrier concentration *n* is calculated using Equation S2.

$$\mu = \frac{\partial I_{SD}}{\partial V_G} \frac{L}{W} \frac{V_{SD}}{\varepsilon_0 \varepsilon_r \, t_{ox}} \qquad (S1)$$

$$n = \frac{\sigma}{e\,\mu} \qquad (S2)$$

Being $\frac{\partial I_{SD}}{\partial V_G}$ the derivation of $I_{SD}$ in FET transfer curves, $V_{SD}$ the source-drain voltage, $\varepsilon_0 \varepsilon_r$ and $t_{ox}$ the permittivity and the thickness (230 nm for interdigitated electrodes, 200 nm for micro-crystal devices) of the dielectric $SiO_x$ layer, respectively, and *e* the elementary charge.

## Temperature dependent conductivity measurements

Figure S6a displays a typical plot of temperature dependent conductivity, as it is also shown in Figure 4d. Figure S6b shows the corresponding Arrhenius plot, where $\ln(\sigma)$ is plotted as a function of $T^{-1}$. Fitting the linear curve yields the activation energy $E_A$.

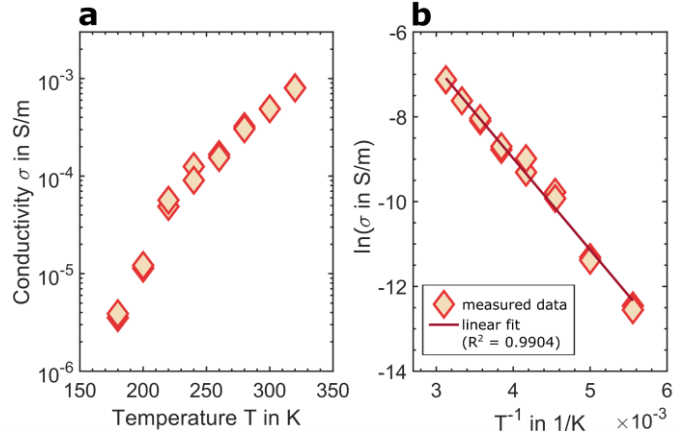

**Figure S6:** Temperature-dependent conductivity measurements. (**a**) Conductivity as a function of temperature of an individual micro-crystal. At every temperature step, several measurements were performed. This Figure corresponds to Figure 4d. (**b**) Arrhenius-plot of the data points shown in (b). The linear fit yields the activation energy $E_A$. The $R^2$ value of 0.9904 indicates the goodness of the linear fitting.

## Calculation of the Coulomb charging energy $E_C$

The estimation of the Coulomb charging energy is performed as described below.

The Coulomb charging energy $E_C$ is given in Equation S3.

$$E_C = \frac{e^2}{2\,C_\Sigma} \qquad (S3)$$

Here, $e$ is the elementary charge and $C_\Sigma$ the total capacitance of the particle to its surrounding. The interparticle capacitance can be estimated using Equation S4.[2,3]

$$C \approx 2\pi\,\varepsilon_0\varepsilon_r r\,\ln\!\left(\frac{r+d}{d}\right) \qquad (S4)$$

Here, $\varepsilon_0$ is the vacuum permittivity, $\varepsilon_r$ the dielectric constant of the surrounding medium (~2.0–2.5 for alkanes and phosphine), $r$ is the NC radius (~0.45 nm) and $2d$ the interparticle distance. Knowing that individual micro-crystals are oriented face-on to the surface, the in-plane Au core-core distance corresponds to the axis *a* and *b* with ~1.9 nm. As the core size is 0.9 nm the interparticle distance is $2d \approx 1.0$ nm. As each NC in the array has eight nearest neighbours, the total capacitance can be calculated to $C_\Sigma = 8\,C$. Accordingly, an estimation of charging energy yields $E_C \approx 276$ meV.

**Details on self-assembly process of Au32-NC micro-crystals**

Different parameters have been investigated to tune the morphology and amount of the micro-crystals. Using a larger quantity of particle solution leads to a higher amount of micro-crystals on the substrate (Figure S7). The same effect can be achieved by increasing the preparation time which allows more crystals to trickle through the subphase. Empirically, 45 min and a volume of 200 µl have shown best results in terms of crystal density along with reasonable preparation times. The influence of the chosen solvent is the following: $Au_{32}$-NC dispersions with hexane yield thicker micro-crystals (up to several hundreds of nanometers), whereas octane lead to the formation of thinner micro-crystals (thicknesses of 50–100 nm). For heptane, intermediate thicknesses can be achieved.

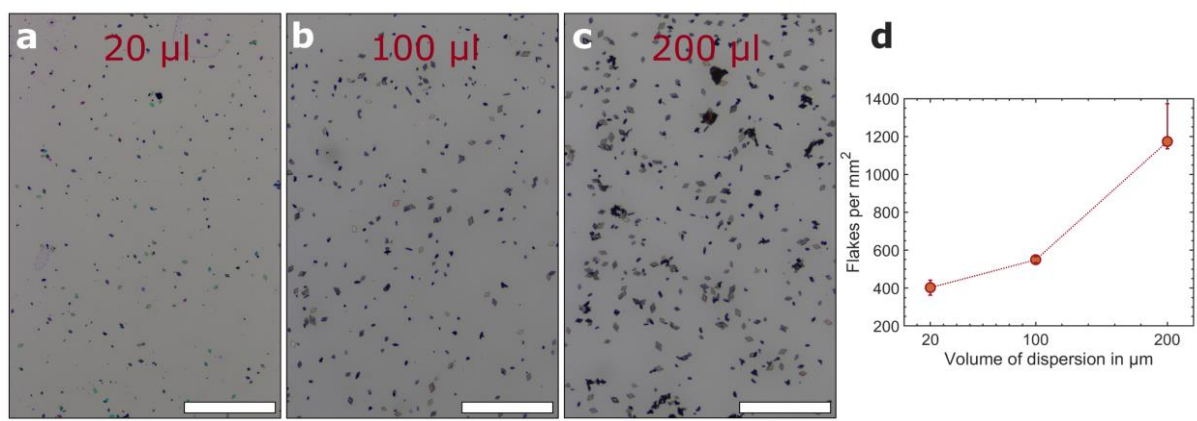

**Figure S7:** Effect of dispersion volume on self-assembly process. At a fixed dispersion solvent (heptane) and concentration (0.5 mM), the number of micro-crystals can be varied. Dispersion volumes of (a) 20 µl, (b) 100 µl and (c) 200 µl were used. (d) Number of micro-crystals per mm² as a function of dispersion volume. Scale bars correspond to 200 µm.

**Mechanism of micro-crystal self-assembly**

To investigate the place and process of the formation of the micro-crystals, a self-built interference reflection microscope was used. A Framos Lt 225 camera with 16 nm/px resolution was used along with a Nikon TIRF objective with oil immersion and a numerical aperture of NA = 1.52. The sample was illuminated by a Rebel High Power LED with a wavelength of 460 nm. A Teflon tube was sealed onto a glass cover slide, filled with acetone and placed onto the microscope. The microscope was focused just above the glass slide into the subphase. An Au$_{32}$-NC solution (0.5 mM, octane) was added onto the subphase to start the process of self-assembly. After the duration of 15 min, the sudden appearance of micro-crystals was observed. From this, we deduce that the process of crystallization does not take place at the substrate but at the liquid-air interface. From there, the micro-crystals start to sink down through the subphase as soon as they reach a critical mass. After reaching the glass slide/substrate, the micro-crystals are able to move laterally within the subphase along the bottom. Upon removal of the subphase, the micro-crystals are deposited onto the substrate.

## Supporting References